%% file: main.tex
\documentclass[10pt,conference]{IEEEtran}
\IEEEoverridecommandlockouts
\usepackage{tabularx}  
\usepackage{graphicx}
\usepackage{url}
\usepackage{hyperref}
\usepackage{listings}
\usepackage{amsmath}
\usepackage{booktabs}
\usepackage{xcolor}
\usepackage{enumitem}

\title{OmniBench-RAG: A Multi-Domain Evaluation Platform for Retrieval-Augmented Generation Tools}

\author{
\IEEEauthorblockN{
    Jiaxuan Liang\textsuperscript{*},
    Shide Zhou\textsuperscript{*},
    and Kailong Wang\textsuperscript{\textdagger}
}
\IEEEauthorblockA{
    Huazhong University of Science and Technology \\
    \{liangjx, shidez, wangkl\}@hust.edu.cn
}
\thanks{\textsuperscript{*}Jiaxuan Liang and Shide Zhou are co-first authors.}
\thanks{\textsuperscript{\textdagger}Kailong Wang is the corresponding author.}
}

\begin{document}
\maketitle

\input{Chapters/abstract}
\input{Chapters/introduction}
\input{Chapters/methodology}
\input{Chapters/evaluation}
\input{Chapters/conclusion}

\bibliographystyle{IEEEtran}
\bibliography{references}

\end{document}

%% file: Chapters/abstract.tex
\begin{abstract}
While Retrieval Augmented Generation (RAG) is now widely adopted to enhance LLMs, evaluating its true performance benefits in a reproducible and interpretable way remains a major hurdle. Existing methods often fall short: they lack domain coverage, employ coarse metrics that miss sub document precision, and fail to capture computational trade offs. Most critically, they provide no standardized framework for comparing RAG effectiveness across different models and domains.

We introduce OmniBench RAG, a novel automated platform for multi domain evaluation of RAG systems. The platform quantifies performance gains across accuracy and efficiency dimensions, spanning nine knowledge fields including culture, geography, and health. We introduce two standardized metrics: Improvements (accuracy gains) and Transformation (efficiency differences between pre RAG and post RAG models), enabling reproducible comparisons across models and tasks. The platform features dynamic test generation, modular evaluation pipelines, and automated knowledge base construction. Our evaluation reveals striking variability in RAG effectiveness, from significant gains in culture to declines in mathematics, highlighting the critical importance of systematic, domain aware assessment.
A demonstration video is available at: \url{https://www.youtube.com/watch?v=BZx83QFcTCI}. Code and datasets: \url{https://github.com/Garnett-Liang/Omnibench-RAG}.
\end{abstract}

%% file: Chapters/introduction.tex
\section{Introduction}
Retrieval-Augmented Generation (RAG) is a key technique for enhancing Large Language Models (LLMs)~\cite{DBLP:conf/nips/LewisPPPKGKLYR020,gao2023retrieval}. By grounding model responses in external, verifiable knowledge, RAG promises to mitigate hallucinations~\cite{zhang2025hallucination}, improve factual accuracy~\cite{li2024enhancing}, and provide up-to-date information~\cite{fan2024survey}. However, the true effectiveness of RAG is far from uniform. Recent studies~\cite{li2025lara, DBLP:conf/acl/NiBGC24} reveal a significant disparity: while RAG can boost the accuracy of smaller models like Llama-3.2-3B-Instruct by as much as 38.12\%, its impact on state-of-the-art models such as GPT-4o, which excel with extended context windows, is often less pronounced. This variability, which depends not only on the model's scale but also heavily on the knowledge domain, underscores a critical challenge: the lack of a systematic platform to quantify the value of RAG across these diverse contexts.

Current RAG evaluation approaches suffer from fundamental limitations that impede reproducible and comprehensive assessment:  
\textbf{First, they lack automated multi-domain evaluation capabilities and rely on non-deterministic components.} Existing benchmarks typically require manual configuration for each knowledge domain and fail to provide unified assessment across diverse fields like finance, healthcare, or culture, making cross-domain performance analysis labor-intensive and inconsistent. Moreover, key metrics in leading frameworks (e.g., LLM-based scoring in Ragas \cite{es2024ragas}) inadvertently introduce randomness due to reliance on large language models in the evaluation loop, undermining result reproducibility.  
\textbf{Second, they employ static datasets and coarse-grained metrics.} Most frameworks rely on fixed benchmarks and document-level retrieval metrics (e.g., MRR@k\cite{DBLP:conf/trec/Voorhees99}), missing the critical sub-document precision needed to assess whether models extract specific facts accurately. They also lack the ability to dynamically generate test cases that probe complex reasoning patterns.  
\textbf{Third, they fail to capture the computational trade-offs inherent in RAG systems.} Without automated profiling of resource utilization and efficiency metrics, practitioners cannot make informed decisions about the cost-benefit trade-offs of deploying RAG in production environments.  

Our work addresses these limitations with \textbf{OmniBench-RAG}, an automated evaluation platform that orchestrates end-to-end assessment of RAG systems across multiple dimensions. The platform introduces several key technical innovations that transform RAG evaluation from manual, ad-hoc testing into systematic, reproducible analysis. At its core, OmniBench-RAG employs an automated parallel evaluation architecture that executes side-by-side comparisons of vanilla and RAG-enhanced models, capturing fine-grained performance metrics including latency, GPU utilization, and memory consumption without manual intervention. This dual-track system ensures that any observed performance differences can be directly attributed to the RAG pipeline's influence.
To enable fair comparisons across heterogeneous models and domains, we introduce a standardized quantification framework with two novel metrics: \textbf{Improvements} for measuring absolute accuracy gains and \textbf{Transformation} for capturing normalized efficiency trade-offs between pre-RAG and post-RAG models.

The platform's modular knowledge base construction pipeline automatically processes domain-specific documents through integrated parsing, chunking, embedding via FAISS\cite{DBLP:journals/tbd/JohnsonDJ21}, and indexing operations. This automation extends to supporting custom document uploads, enabling personalized RAG evaluation scenarios without requiring technical expertise in vector database management. By automating the entire evaluation workflow~(i.e., from dataset generation to multi-dimensional analysis across nine knowledge domains), OmniBench-RAG provides actionable insights into the nuanced interplay between model architecture, retrieval quality, and computational efficiency, empowering both researchers and practitioners to make data-driven decisions about RAG deployment.

\textbf{The main contributions of our evaluation platform are:}
\begin{itemize}[leftmargin=*]
\item \textbf{Automated Parallel Evaluation Architecture.} Dual-track system that simultaneously evaluates vanilla and RAG-enhanced models, automatically capturing accuracy, latency, GPU utilization, and memory consumption metrics for direct performance comparison.

\item \textbf{Standardized Quantification Metrics.} Introduction of \emph{Improvements} (absolute accuracy gains) and \emph{Transformation} (normalized efficiency trade-offs) metrics for reproducible cross-model and cross-domain comparisons.

\item \textbf{Modular Knowledge Base Construction Pipeline.} Automated document processing system with integrated parsing, chunking, FAISS embedding, and indexing, supporting custom document uploads for domain-specific RAG evaluation.
\end{itemize}

%% file: Chapters/methodology.tex
\section{OmniBench-RAG}

\begin{figure}[htbp]
    \centering\vspace{-0.5cm}
    \includegraphics[width=0.8\columnwidth]{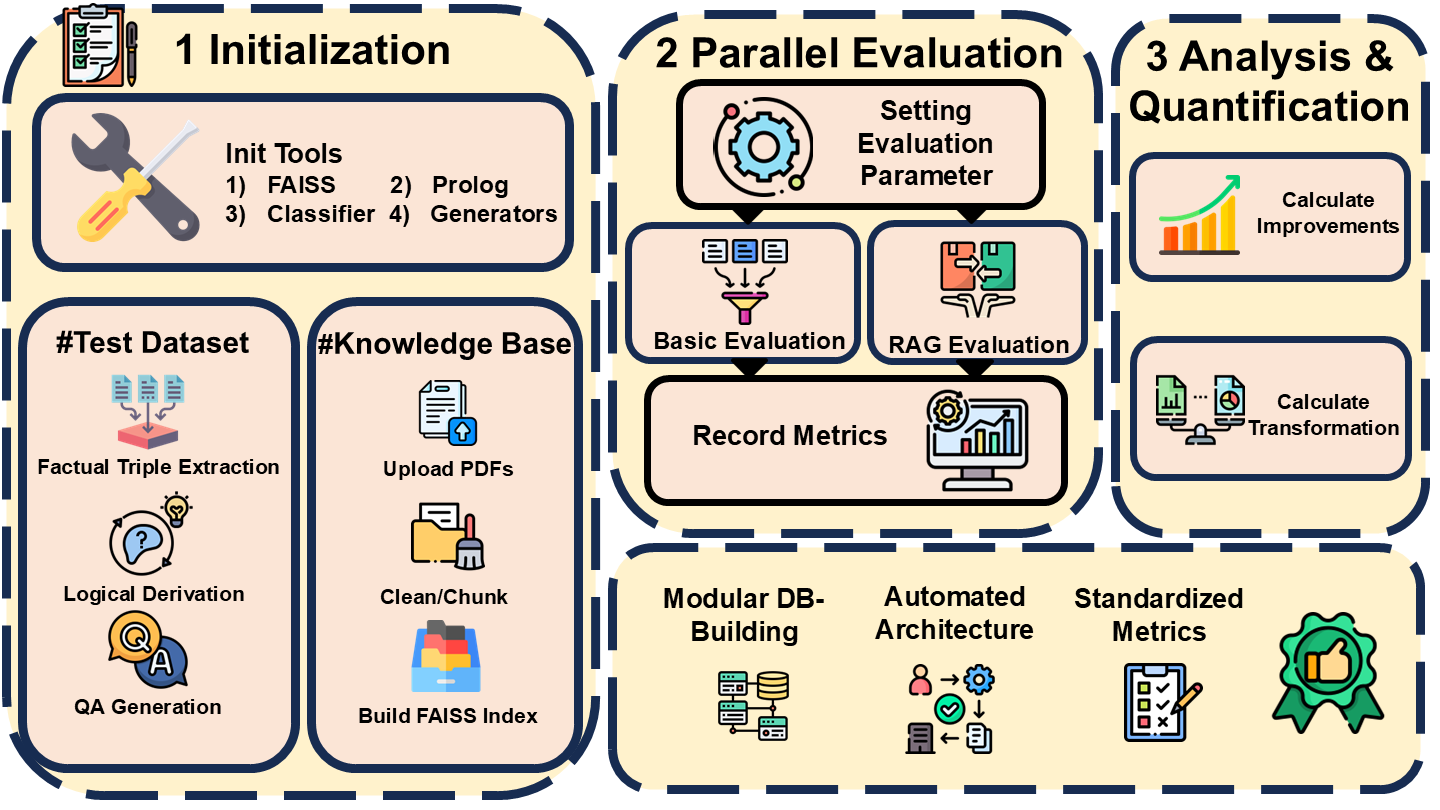} \vspace{-0.5cm}
    \caption{Workflow of OmniBench-RAG}
    \label{fig:workflow} \vspace{-0.3cm}
\end{figure}

The cornerstone of OmniBench-RAG is a systematic, side-by-side evaluation workflow designed to isolate and quantify the multi-dimensional impact of Retrieval-Augmented Generation. As illustrated in Figure~\ref{fig:workflow}, the platform executes two parallel evaluation tracks---a \textbf{Basic Evaluation} for the vanilla model and a \textbf{RAG Evaluation} for the retrieval-enhanced model---using identical test datasets. This dual-track process is essential for calculating our standardized enhancement metrics for both Improvements and Transformation. The entire workflow is orchestrated through three primary stages, which we detail in the following sections.

\subsection{Stage 1: Initialization and Asset Preparation}
The initial stage prepares all necessary assets for the evaluation, ensuring a consistent and reproducible testing environment. The process commences with \textbf{System Configuration}, where the user loads the target language models and initializes core platform components, including the FAISS (Facebook AI Similarity Search) library for vector retrieval and the dynamic dataset generation engine.

A core component of this stage is the \textbf{Knowledge Base Construction} for the \textbf{RAG Evaluation} track. A user provides a corpus of domain-specific documents (e.g., in PDF format) that serves as the external knowledge source. The platform then initiates an automated pipeline: documents are parsed, cleaned, and partitioned into smaller, semantically coherent passages or \textit{chunks}. Each chunk is subsequently encoded into a high-dimensional vector using a pre-selected embedding model and indexed within a FAISS instance. This procedure results in a highly efficient, searchable knowledge base, which is foundational for the retrieval operations in the subsequent stage.

Finally, this stage prepares the \textbf{Test Dataset}, a critical shared asset for both evaluation tracks. OmniBench-RAG offers two methods for dataset preparation:
\begin{itemize}[leftmargin=*]
    \item \textbf{Loading an Existing Dataset:} Users can directly load a pre-existing question-answering (QA) dataset, typically in a standardized JSON format, allowing for evaluation on established benchmarks.
    \item \textbf{Dynamic Dataset Generation:} To probe model capabilities with novel and complex queries, the platform integrates a dynamic generation engine. It adopts a logic-based methodology similar to that of Drowzee \cite{10.1145/3689776} to automatically produce novel test cases. This process unfolds in three steps:
    \begin{enumerate}[leftmargin=*]
        \item \textit{Factual Triple Extraction:} The engine first extracts factual triples from knowledge sources like Wikipedia to act as seed knowledge.
        \item \textit{Logical Derivation:} It then applies a set of logical inference rules to these initial facts to derive new, more complex knowledge that is not explicitly stated in the source text.
        \item \textit{Question-Answer Generation:} Finally, these derived facts are programmatically converted into challenging QA pairs using predefined templates, ensuring a continuous supply of diverse and logically sophisticated test cases.
    \end{enumerate}
\end{itemize}

\subsection{Stage 2: Parallel Evaluation Execution}
With all assets prepared, the platform proceeds to the core evaluation stage, which is defined by the parallel execution of the \textbf{Basic Evaluation} and \textbf{RAG Evaluation} tracks. This controlled comparison is critical for isolating the effects of the RAG pipeline.In the \textbf{Basic Evaluation} track, the model operates in a standard, non-retrieval mode. Presented with a question from the test set, it generates an answer based solely on its internal, pre-trained knowledge. During this process, the platform leverages a binary question-answering dataset \cite{10.1145/3689776} to systematically evaluate performance. The platform meticulously records a set of fundamental performance metrics, which we designate as the \textit{Base Metrics}:
\begin{itemize}[leftmargin=*]
\item \textbf{Accuracy Score (\(S_{\text{base}}\)):} The correctness score of the model's answer is determined by comparing its response to ground truth labels using the BinaryAnswerClassifier, a DistilBERT-based binary classifier fine-tuned on question-answering datasets. This classifier outputs a binary judgment (positive/negative) aligned with the dataset's gold-standard answers.
\item \textbf{Response Time (\(T_{\text{base}}\)):} Measured as the latency from receiving the question to producing the complete answer, captured using high-resolution timers.
\item \textbf{GPU Utilization (\(U_{\text{gpu\_base}}\)):} The peak GPU memory consumption during the inference process, recorded using system monitoring tools.
\item \textbf{Memory Utilization (\(U_{\text{mem\_base}}\)):} The peak system memory (RAM) consumption during the inference process, tracked via OS-level memory profiling.
\end{itemize}To isolate the impact of retrieval, the \textbf{RAG Evaluation} track subjects the same model to an identical query, but with the augmentation process enabled. This process involves retrieving the Top-K most semantically relevant knowledge chunks from the FAISS index and augmenting the original prompt with this information before generating an answer. The platform applies the same measurement methodologies as in the baseline track to compute the \textit{RAG Metrics}: \(S_{\text{RAG}}\), \(T_{\text{RAG}}\), \(U_{\text{gpu\_RAG}}\), and \(U_{\text{mem\_RAG}}\). This parallel execution ensures that any observed performance difference between the two tracks can be directly attributed to the influence of the RAG pipeline.

\subsection{Stage 3: Comparative Analysis and Quantification}
The final stage synthesizes the raw data collected from the two parallel tracks to produce a holistic, quantitative assessment of the RAG system's value. This is achieved by calculating two distinct categories of metrics: Improvements and Transformation.

To quantify the improvement, we introduce two standardized metrics. The first, \textbf{Improvements}, measures the absolute improvement provided by RAG and is formally defined as:
\begin{equation}
    Improvements = S_{\text{RAG}} - S_{\text{base}}
\end{equation}
The second, \textbf{Transformation}, provides a comprehensive measure of the 
efficiency trade-offs introduced by the RAG pipeline. It aggregates normalized 
ratios across three key performance dimensions: \textit{response time}, 
\textit{GPU memory utilization}, and \textit{system memory consumption}. 
Formally, it is defined as:

\begin{equation}
    Transformation = \frac{w_{\text{time}}}{r_{\text{time}}} + 
                            \frac{w_{\text{gpu}}}{r_{\text{gpu}}} + 
                            \frac{w_{\text{mem}}}{r_{\text{mem}}}
\end{equation}

where:
\begin{itemize}
    \item $r_{\text{time}} = \frac{T_{\text{RAG}}}{T_{\text{base}}}$, 
          $r_{\text{gpu}} = \frac{U_{\text{gpu\_RAG}}}{U_{\text{gpu\_base}}}$, 
          $r_{\text{mem}} = \frac{U_{\text{mem\_RAG}}}{U_{\text{mem\_base}}}$ 
          denote the response time ratio, GPU memory ratio, and system memory ratio, respectively;
    \item $w_{\text{time}}$, $w_{\text{gpu}}$, and $w_{\text{mem}}$ are the weights assigned to each efficiency dimension. In OmniBench-RAG, these weights are user-configurable to reflect different deployment priorities. For our experiments, we use the default values of $w_{\text{time}}=0.4$, $w_{\text{gpu}}=0.3$, and $w_{\text{mem}}=0.3$ to represent a typical interactive application scenario.
\end{itemize}

The weighted aggregation in \textit{Transformation} reflects deployment priorities, where latency dominates real-time applications, while memory constraints drive edge deployments.

A \textit{Transformation} score greater than 1.0 indicates that the RAG-enhanced model 
achieves better efficiency (lower resource consumption) compared to the baseline, 
while a score less than 1.0 suggests increased overhead. This metric enables fair 
comparisons across heterogeneous models and domains by accounting for both 
accuracy improvements (\textit{Improvements}) and efficiency trade-offs.

Culminating the analysis, OmniBench-RAG aggregates these results into comprehensive reports. A key feature of the platform is its ability to generate domain-specific breakdowns, allowing for a granular analysis of RAG performance across nine distinct knowledge fields: Geography, History, Health, Mathematics, Nature, People, Society, Technology, and Culture. This multi-faceted output provides researchers with a deep understanding of the crucial trade-offs between accuracy gains and computational costs inherent to RAG systems.

%% file: Chapters/evaluation.tex
\section{Evaluation}

To assess the practical effectiveness of OmniBench-RAG, we conducted a comprehensive evaluation of the Qwen model across nine distinct knowledge domains: culture, geography, history, health, mathematics, nature, people, society, and technology. Each domain includes queries generated from platform-built-in datasets and Wikipedia content via dynamic generation methods, incorporating diverse reasoning patterns (e.g., negation, inverse, complex) ~\cite{10.1145/3689776}, and every model variant was evaluated both with and without RAG enhancement.

Our results reveal significant variability in RAG’s impact across domains, with notable improvements in several areas. The most substantial gains were observed in culture (+17.1\%), people (+16.7\%), nature (+11.7\%), and technology (+10.7\%)—domains where RAG augmentation consistently enhanced performance. A key driver behind these gains appears to be the strong alignment between RAG source materials and domain-specific query requirements. For all domains, RAG materials primarily consist of dozens of relevant papers retrieved from arXiv, with full references accessible via our code repository. For instance, the culture domain leveraged datasets such as Reviving Cultural Heritage: A Novel Approach for Comprehensive Historical Document Restoration ~\cite{zhang2025reviving}, which is rich in narrative and contextual knowledge that aligns naturally with QA tasks. Such materials, characterized by descriptive, conceptually dense content, provide the model with precise, relevant context to ground its answers—directly contributing to the observed accuracy gains.
In contrast, domains such as health (-18.3\%) and math (-25.6\%) saw declines, probably due to misalignment between retrieval materials and domain-structured rule-based reasoning demands. This also correlates with structural mismatches: rule-based domains require precise symbolic retrieval, whereas generic chunking fails to preserve logical dependencies. This aligns with findings in ~\cite{mohapatra2021domain}, where rule-based domains demand precise symbolic retrieval and generic chunking fails to preserve logical dependencies by conflating relevant and irrelevant tokens.

To further explore how retrieval content quality shapes performance, we utilized OmniBench-RAG with a custom document upload feature. This allows users to experiment with domain-specific materials, enabling personalized investigation into how different datasets influence RAG effectiveness—empowering deeper exploration of the relationship between source content and model performance.

The Transformation metric, which captures efficiency trade-offs, shows that RAG generally introduces moderate overhead (values $<$ 1) due to retrieval and context-processing costs, consistent with expectations. Notably, math (1.1181) is an exception: its efficiency gain, despite accuracy drops, likely stems from irrelevant retrievals reducing the model’s need for complex reasoning, highlighting the critical role of high-quality source materials.

\begin{table}[htbp] 
\centering
\caption{Qwen Performance Across Nine Knowledge Domains}  
\label{tab:qwen_results}  
\begin{tabular}{l|cccc} 
\hline
\textbf{Domain} & \textbf{Baseline} & \textbf{RAG} & \textbf{Improvements} & \textbf{Transformation} \\ 
\hline
Culture        & 51.1\%  & 68.2\%  & +17.1\%  & 0.6403  \\  
Geography      & 78.0\%  & 79.2\%  & +1.3\%   & 0.7497  \\
History        & 42.5\%  & 45.1\%  & +2.6\%   & 0.9911  \\
Health         & 70.0\%  & 51.7\%  & -18.3\%  & 0.8226  \\
Math           & 76.9\%  & 51.3\%  & -25.6\%  & 1.1181  \\
Nature         & 53.3\%  & 65.1\%  & +11.7\%  & 0.7255  \\
People         & 41.6\%  & 58.3\%  & +16.7\%  & 0.8746  \\  
Society        & 67.3\%  & 69.1\%  & +1.8\%   & 0.8354  \\
Technology     & 57.1\%  & 67.9\%  & +10.7\%  & 0.9238  \\
\hline
\end{tabular}
\end{table}

Table~\ref{tab:qwen_results} summarizes Qwen's performance in all nine domains. While Improvements reflect the accuracy gain, the Transformation metric captures the relative efficiency differences between pre-RAG and post-RAG models, offering a fair comparison across knowledge areas with differing task characteristics.
\begin{figure}[htbp]
    \centering
    \includegraphics[width=0.8\columnwidth]{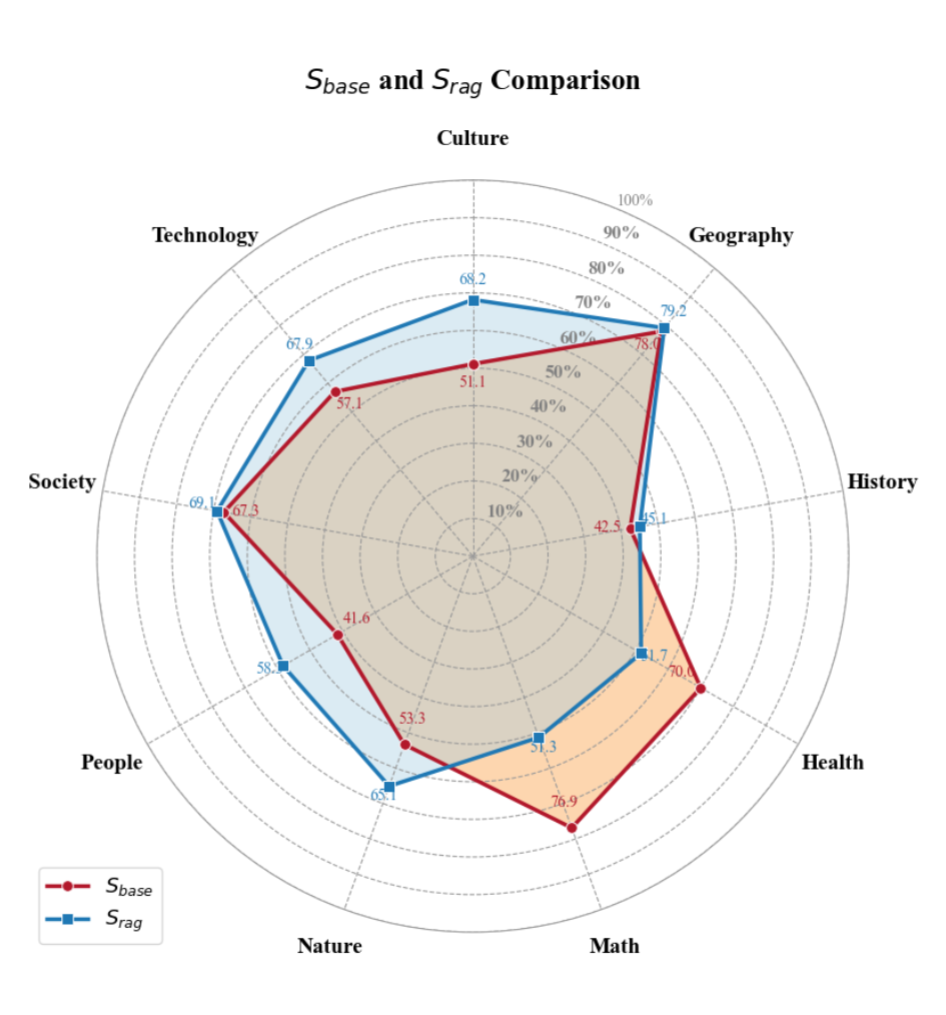}\vspace{-0.3cm}
    \caption{Radar plot of Qwen's accuracy across nine knowledge domains, comparing $S_{base}$ and $S_{rag}$.}\vspace{-0.3cm}
    \label{fig:qwen_radar}
\end{figure}

%% file: Chapters/conclusion.tex
\section{Conclusion}
In conclusion, OmniBench RAG addresses critical gaps in RAG evaluation through its automated, multi domain assessment framework spanning nine knowledge fields. The platform's key innovations, including automated parallel evaluation, dynamic test generation, and standardized metrics, transform RAG assessment from ad hoc testing into systematic, reproducible analysis. Our evaluation reveals striking domain dependent variability, from significant gains in culture and people to unexpected declines in health and mathematics, demonstrating that RAG effectiveness cannot be assumed uniform across domains. By providing automated knowledge base construction and comprehensive performance profiling without manual intervention, OmniBench RAG empowers researchers and practitioners to make data driven decisions about RAG deployment, establishing a new standard for fair and reproducible RAG evaluation across heterogeneous models and domains.